\documentclass[reprint,amsmath,amssymb,aps,prl,superscriptaddress,nofootinbib]{revtex4-2}
\usepackage{graphicx}
\usepackage{dcolumn}
\usepackage{bm}
\usepackage{mathrsfs}
\usepackage{amsmath}
\usepackage{hyperref}
\usepackage{latexsym}
\usepackage{array}
\usepackage{color}
\usepackage{multirow}

\hypersetup{hypertex=true, colorlinks=true, linkcolor=blue, urlcolor=blue, citecolor=magenta}

\begin{document}

\title{Stochastic Model and Optimal Control of an Active Tracking Particle with Information Processing}

\author{Tai Han}
\affiliation{Institute of Theoretical Physics, Chinese Academy of Sciences, Beijing 100190, China}
\affiliation{School of Physical Sciences, University of Chinese Academy of Sciences, Beijing 100049, China}

\author{Fanlong Meng}
\email{fanlong.meng@itp.ac.cn}
\affiliation{Institute of Theoretical Physics, Chinese Academy of Sciences, Beijing 100190, China}
\affiliation{School of Physical Sciences, University of Chinese Academy of Sciences, Beijing 100049, China}
\affiliation{Wenzhou Institute, University of Chinese Academy of Sciences, Wenzhou, Zhejiang 325000, China}

\begin{abstract}
Living systems often function with regulatory interactions, but the question of how activity, stochasticity and regulations work together for achieving different goals still remains puzzling.
We propose a stochastic model of an active tracking particle with information processing, where the entropy production and information flow are discussed, with the generalised fluctuation theorem serving as a benchmark for verifying the probability setups.
Based on the model, the system performance, in terms of the first passage steps and the total energy consumption, are analysed in the variable space of (measurement error, control field), leading to discussions on optimal controls of the system.
Not only elucidating the basic concepts involved in a stochastic active system with information processing, this prototypical model could also inspire more elaborated modelings of natural smart organisms and industrial designs of controllable active systems with desired physical performances in the future.
\end{abstract}

\maketitle

Among the theoretical concepts proposed for understanding \emph{physics of life},
active matter can be regarded as a very successful trial~\cite{ramaswamy2010mechanics,marchetti2013hydrodynamics,tailleur2022,te2025metareview}.
Not only now serving as a model system for studying the far-from-equilibrium statistical physics,
active matter, which is often assumed with simple physical rules, also keeps refreshing our understandings of the living systems, in terms of both the individual behaviours such as microorganism locomotion~\cite{elgeti2015physics,berg1973bacteria,bayly2011propulsive,Hu2024} and the collective responses such as bird flocking~\cite{vicsek1995,toner1995,cavagna2014bird,ginelli2016physics}.
Although different active models have gained great successes in analysing the physical responses in customised cases, such models are usually oversimplified for describing the biological activities of most creatures with complicated regulatory interactions, which often include signal sensing, decision making and adaptive responses~\cite{pardee2006regulatory,bich2016biological,bashor2018understanding,Levine2023}.

In living systems, the acquired knowledge of their own internal status or the external environment can be utilised to modulate their responses, i.e., regulations concerning information flow, and such regulatory examples range from the microscopic ones such as algae phototaxis~\cite{foster1980light,witman1993chlamydomonas} to the macroscopic ones such as human crowding~\cite{schmidt1979human,corbetta2023physics}.
Inspired by such examples,
there have been various studies which implement external controls to achieve the desired physical performances of soft or active matter systems~\cite{McDonald2023,Takatori2025,Alvarado2025,Gompper2025}, e.g., passive-to-active transformation of a Brownian particle~\cite{Huang2020,Baldovin2023}, microswimmer navigation with minimum time or energy consumption~\cite{Mano2017,Selmke2018, Liebchen2019,Schneider2019,Daddi-Moussa-Ider2021,Monthiller2022,Piro2022,Nasiri2023}, and maximum work efficiency of information engines~\cite{malgaretti2022,Saha2023,Cocconi2024}.
However, despite these initial attempts illustrating the importance of regulations in controlling soft or active matter systems, how to formulate the information flow, entropy production and physical performances in a unified theoretical framework still remains questionable, rendering many difficulties in understanding the responses of a regulated living system from a physical perspective.

\begin{figure}[htbp]
\includegraphics[width=0.4378\textwidth]{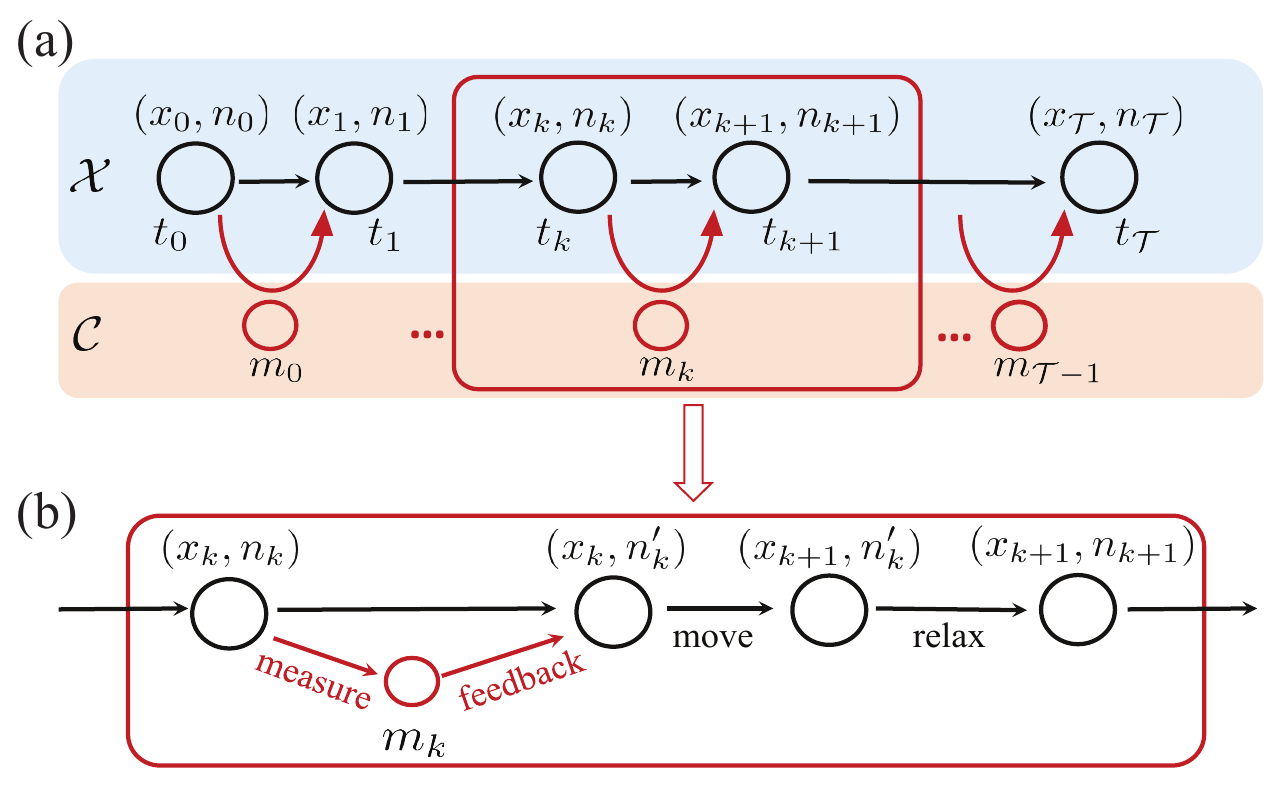}
\caption{(a) Evolution of the system consisting of a physical subsystem $\mathcal{X}$ in charge of the particle locomotion and a control (measurement - feedback) subsystem $\mathcal{C}$ in charge of the information-based regulation, and (b) zoom-in of the system dynamics during the time step, $t_{k}\rightarrow t_{k+1}$. }\label{F1}
\end{figure}

In this work,
we propose a stochastic model of an active tracking particle with information processing, based on the Bayesian description.
We discuss about the entropy production and the information flow of the system, where the generalised fluctuation theorem serves as a benchmark for verifying the probability setups.
With the model, we investigate how the system performance, in terms of the first passage steps and the total energy consumption, can depend on the information-related processes including measurements and feedback controls, providing the optimal control protocols in the corresponding parameter space.

\paragraph{System setup.}
For seeking the transparency of the model,
we take a one-dimensional and descritised description to illustrate how a regulated active particle should move, as shown in Fig.~\ref{F1}.
The state of the active particle at any time, $t_{k \in \mathbb{Z}}$, is described by two variables $(x_{k}, n_{k})$, where $x_{k}=s~\! \Delta x$ denotes its location with $s\in \mathbb{Z}$ and $\Delta x$ as the step length, and $n_{k}\in (L, R)$ denotes its orientation pointing to the left or the right; without the information processing, the model reduces to the run-and-tumble model describing the dynamics of a bacterium~\cite{berg2004coli,tailleur2008statistical}.
The regulatory purpose of the active particle is to move from the location $x=0$ to the destination $x=\mathcal{D}\Delta x$ along a straight line without any step moving to the left, i.e., the location should be $x=d~\!\Delta x$ at any time $t_{d}$, and this is a typical tracking problem in control theories.
To achieve feedback controls, we consider the particle carrying a magnetic moment of magnitude $m$ in the same direction as its orientation, and the orientation can be controlled by applying an external magnetic field; such a model can be easily generalised to other controlling cases, e.g., by utilising light signal on algae~\cite{foster1980light,witman1993chlamydomonas} or chemical concentration on bacteria~\cite{Celani2010,Nava2018}.
As shown in Fig.~\ref{F1}, the particle starts moving from state $(x_{0}, n_{0})$ at time $t_{0}$, and after measurement - feedback control- active motion - relaxation, the particle reaches a new state $(x_{1}, n_{1})$ at time $t_{1}$.
By repeating the above processes, the particle stops moving after arriving at the destination $(\mathcal{D} \Delta x, n_{\mathcal{T}})$ after $\mathcal{T}$ time steps ($\mathcal{T}\geq \mathcal{D}$);
note that the particle motion can deviate from the desired trajectory (keep moving to the right without any step moving to the left) due to the intrinsic stochasticity of the system.

We utilise a bipartite Bayesian network~\cite{Ito2016,Hartich2014} to describe the dynamics of the active particle, consisting of a physical subsystem $\mathcal{X}$ in charge of the particle locomotion as an ordinary run-and-tumble particle and a control (measurement - feedback) subsystem $\mathcal{C}$ in charge of the information-based regulation.
Note that the system ($\mathcal{X}$ + $\mathcal{C}$) is placed in a heat bath, subjected to thermal fluctuations.
With this treatment, the regulation of the whole process simply follows the classical Bellman description~\cite{bellman1954theory} of dynamic programming where the total optimal cost is the summation of all the costs at every time step, so it is sufficient to focus on the regulation process during a single time step, $t_{k}\rightarrow t_{k+1}$ with $0\leq k<\mathcal{T}$.
For achieving regulatory purpose during any arbitrary time step, $t_{k}\rightarrow t_{k+1}$, we take the Maxwell-demon-like setup of the measurement - feedback process~\cite{maxwell2012theory, szilard1964decrease,szilard1929entropieverminderung}, and it is detailed as follows.
At time $t_{k}$, the state of the particle is $(x_{k}, n_{k})$,
where $n_{k}=L, R$ with equal probability at equilibrium, i.e., $p(n_{k}=L, R)=p_{\mathrm{eq}}^{0}(n_{k}=L, R)=1/2$.
Then we measure its orientation denoted as $m_{k}$, and such measurements have an error (probability for a wrong measurement)~\cite{sagawa2010generalized,v1_brillouin2013science, v1_Lan2012}, $0 \leq\epsilon\leq1$, which may arise from limited instrument precision, thermal fluctuation, etc.; in other words, we can define a conditional probability characterising this measurement: $p(m_{k}|n_{k})=1-\epsilon$ for $m_{k}=n_{k}$ and $p(m_{k}|n_{k})=\epsilon$ for  $m_{k}\neq n_{k}$.
If the measured orientation is pointing to the left, i.e., $m_{k}=L$, then we apply a magnetic field of the magnitude $B$ and the direction pointing to the right during a finite time interval $\Delta t$ to change its orientation;
if the measured orientation is pointing to the right, i.e., $m_{k}=R$, then we do not apply the magnetic field and let the particle orientation freely evolve during $\Delta t$ driven by thermal fluctuation.
We use $n_{k}'$ to denote the particle orientation after the measurement and feedback control.
The transition, $n_{k}\rightarrow n_{k}'$, which is critical in the regulatory process, includes the nodes, $n_{k}$, $m_{k}$, and $n_{k}'$ as shown in Fig.~\ref{F1}, and the transition can be described by the conditional probability, $p(n'_{k}|n_{k},m_{k})$, with its explicit expressions shown in Table~\ref{t1} of the End Matter, by mapping this dynamic process to a typical two-state Kramers process~\cite{supplement,Kramers,hanggi1990reaction}.
For example, no magnetic field is applied in the case of ($n_{k}=L$,  $m_{k}=R$),  where the particle orientation will evolve freely during $\Delta t$, and we can obtain the probabilities of finding the particle
with $n_{k}'=L$ and $n_{k}'=R$ as,  $p_{\mathrm{eq}}^{0}(n_{k}'=L)+[1-p_{\mathrm{eq}}^{0}(n_{k}'=L)]\cdot \exp[-k_{0}\Delta t]$ and $p_{\mathrm{eq}}^{0}(n_{k}'=R)-[1-p_{\mathrm{eq}}^{0}(n_{k}'=L)]\cdot \exp[-k_{0}\Delta t]$, respectively, where the transition rate is $k_{0}=k_{0}^{L\rightarrow R}+k_{0}^{R\rightarrow L}=2k_{0}^{L\rightarrow R}$ with $k_{0}^{L\rightarrow R}$ and $k_{0}^{R\rightarrow L}$ as the transition rate from $n_{k}=L$ to $n_{k}'=R$ and that from $n_{k}=R$ to $n_{k}'=L$ without magnetic field, respectively.
In following discussions, the notation
$u_{p}=\exp[-k_{0}\Delta t]$ (related with unchanged population) is taken for simplicity.
When applying magnetic field, the probability, $p(n'_{k}|n_{k},m_{k})$, depends explicitly on the strength of the magnetic field due to thermal fluctuations, as shown Table~\ref{t1}.
After the measurement - feedback process, we let the particle move along its regulated orientation with a constant step length, $\Delta x$, which is described by the process $(x_{k}, n_{k}')\rightarrow(x_{k+1}, n_{k}')$, i.e., for the `move' process (analogous to the `run' step of a run-and-tumble bacterium) in Fig.~1(b), the spatial position of the particle will become $x_{k+1}= x_{k}+\Delta x$ for $n'_{k} = R$, and become $x_{k+1}= x_{k}-\Delta x$ for $n'_{k} = L$. Such dynamic process can be achieved in various active systems, e.g., by applying magnetic field on platinum-nickel-gold nanorods~\cite{v1_Kline2005}, shining light on gold-polystyrene Janus particles~\cite{v1_Qian2013} or applying electric field on metal-dielectric Janus particles~\cite{yan2016reconfiguring,boymelgreen2022role}.
After this, we let the particle orientation relax to the equilibrium state with its orientation distribution as $p_{\mathrm{eq}}^{0}(n_{k+1}=L, R)=1/2$ at time $t_{k+1}$.
For those familiar with bacteria swimming in the run-and-tumble mode, the above setup can be treated as a bacterium moving with a `brain' which performs measurement and feedback control.
By repeating the above processes from $t_{0}$ to $t_{\mathcal{T}}$, the particle can arrive at its destination after $\mathcal{T}$ time steps, with its state as $(\mathcal{D}\Delta x, n_{\mathcal{T}})$.

\paragraph{Entropy production and information flow.} The total entropy production in the whole system consists of the entropy production in both the subsystem $\mathcal{X}$ and its connected heat bath \cite{Ito2016, sagawa2013role}, i.e., $\Delta\mathcal{S}_{\mathrm{tot}}=\Delta\mathcal{S}_{\mathcal{X}}+\Delta\mathcal{S}_{\mathrm{bath}}=\sum_{k=0}^{\mathcal{T}-1}(\Delta s_{\mathcal{X}}^k+\Delta s_{\mathrm{bath}}^{k})$, where $\Delta s_{\mathcal{X}}^k$ and $\Delta s_{\mathrm{bath}}^k$ denote the entropy production of the subsystem $\mathcal{X}$ and the heat bath during the time step, $t_{k}\rightarrow t_{k+1}$, respectively.
The entropy production of the subsystem $\mathcal{X}$, $\Delta s_{\mathcal{X}}^k=\ln[ p(n_{k})/p(n_{k+1})]=0$, since particle orientation at time $t_{k}$ and $t_{k+1}$ are both in the equilibrium state, i.e., $p(n_{k})=p(n_{k+1})=1/2$.
The entropy change of the heat bath, $\Delta s_{\mathrm{bath}}^{k}$, includes two contributions, $\Delta s_{\mathrm{bath}}^{k1}$ during the `evolution' process of $(x_{k},n_{k})\rightarrow (x_{k},n_{k}')$, and $\Delta s_{\mathrm{bath}}^{k2}$ during the `relax' process of $(x_{k+1},n_{k}')\rightarrow(x_{k+1},n_{k+1})$.
The first contribution is $\Delta s_{\mathrm{bath}}^{k1}=\ln [p(n'_{k}|n_{k},m_{k})/p_{B}(\hat{n}_{k}|\hat{n}'_{k},\hat{m}_{k})]$,
where $p(n'_{k}|n_{k},m_{k})$ denotes the transitional probability as introduced above, and $p_{B}(\hat{n}_{k}|\hat{n}'_{k},\hat{m}_{k})$ denotes the backward transition probability from $\hat{n}'_{k}$ to $\hat{n}_{k}$ depending on the memory $\hat{m}_{k}$, with $\hat{n}_{k}$, $\hat{n}'_{k}$ and $\hat{m}_{k}$ as time-reversed $n_{k}$, $n'_{k}$ and $m_{k}$, respectively.
The second contribution is
$\Delta s_{\mathrm{bath}}^{k2}=\ln [p(n_{k+1}|n'_{k})/p_{B}(\hat{n}'_{k}|\hat{n}_{k+1})]$, denoting the relaxation process of the particle orientation, and it is always positive.
For explicit expressions of the above mentioned probabilities, one can refer to the supplemental materials for more details~\cite{supplement}.
Note that the entropy production during the `move' process is zero, i.e., the `move' process is reversible by reversing the direction of the particle, which is different from the case of a particle driven by a constant force without changing the driving direction~\cite{supplement}.

For characterising how measurement and feedback control can influence the dynamics of the active particle, we apply the information quantity,  $\Theta_{k} =I_{k}^{\mathrm{fin}}-I_{k}^{\mathrm{tr}}-I_{k}^{\mathrm{ini}}$~\cite{Ito2016},
where the mutual information $I_{k}^{\mathrm{fin}} =I(n'_{k}:m_{k})=\ln [p(n'_{k},m_{k})/p(n'_{k})p(m_{k})]$ characterises the correlation between $n'_{k}$ and $m_{k}$, the mutual information  $I_{k}^{\mathrm{tr}}=I(m_{k}:n_{k})=\ln [p(m_{k},n_{k})/p(m_{k})p(n_{k})]$ denotes the transfer entropy from the physical system $\mathcal{X}$ to the control system $\mathcal{C}$, and $I_{k}^{\mathrm{ini}}=0$ denotes no correlation between $n_{k}$ and previous measurements, $m_{l<k}$.
By summing them up, we obtain the dynamic information flow in the system, $\Theta_{d}= \sum_{k=0}^{\mathcal{T}-1}\Theta_{k}$.
One can prove that the total entropy production, $\Delta\mathcal{S}_{\mathrm{tot}}$, and the dynamic information flow, $\Theta_{d}$, satisfy the generalised integral fluctuation theorem: $
\langle e^{-\Delta\mathcal{S}_{\mathrm{tot}}+\Theta_{d} }\rangle=1$,
as obtained in other information-incorporated stochastic systems~\cite{Sagawa2010,Toyabe2010,v1_Munakata2012,Sagawa2012,v1_Ito2013,Ito2016,v1_Zeng2021},
which is utilised as a benchmark in this work for verifying the probability setup of the above dynamic processes.

\begin{figure}[htbp]
\includegraphics[width=0.4178\textwidth]{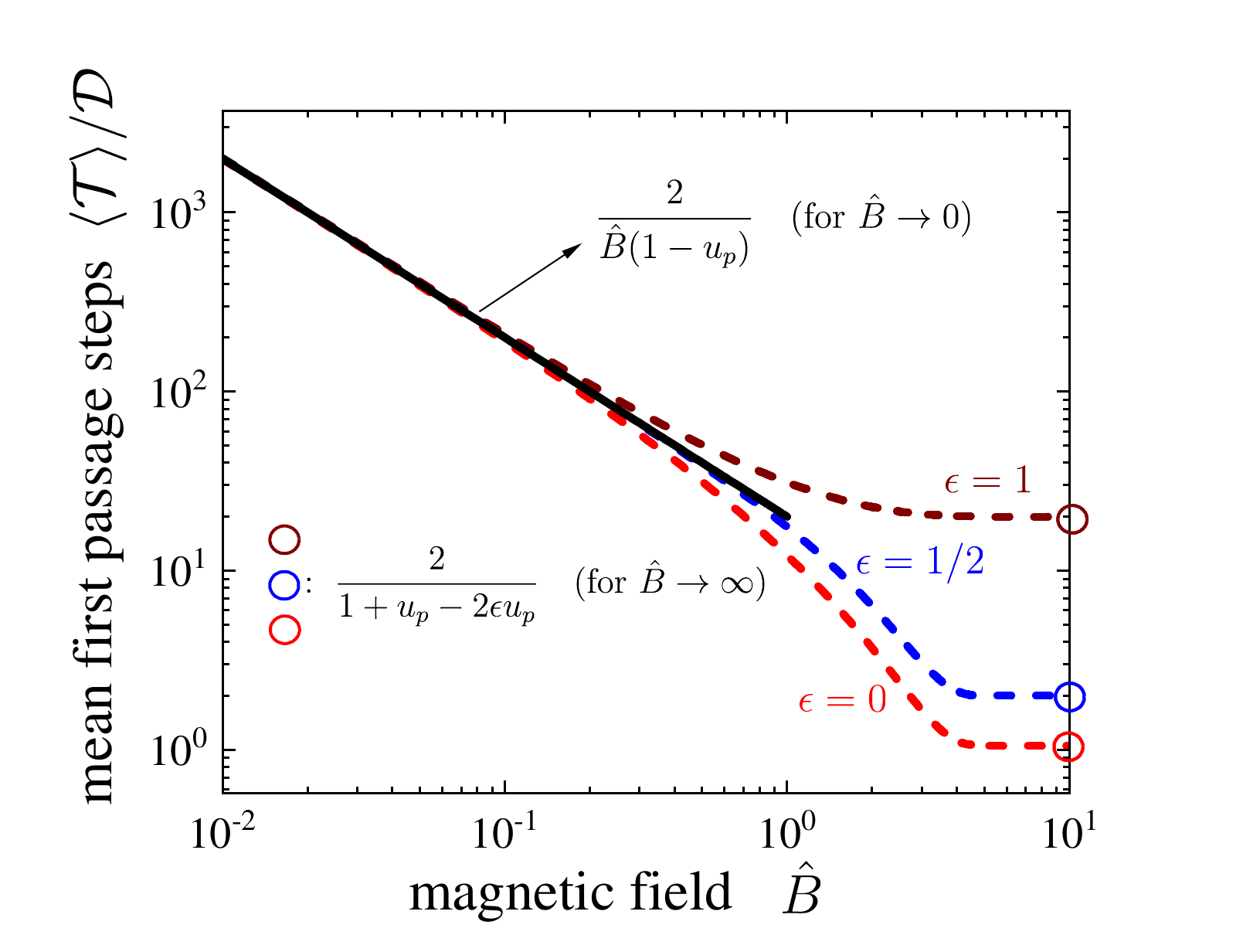}
\caption{Mean first passage steps $\langle \mathcal{T}\rangle$ as a function of the magnetic field $\hat{B}$, for $\epsilon=0$ (red), 1/2 (blue) and 1 (brown). The black solid line is obtained by $\langle \mathcal{T}\rangle /\mathcal{D} \simeq 2/[\hat{B} (1 -u_{p})]$ for $\hat{B}\rightarrow0$, and the round circles for different $\epsilon$ denote the values obtained by $\langle \mathcal{T}\rangle/\mathcal{D} \simeq2/(1+u_{p}-2 \epsilon u_{p})$ for $\hat{B}\rightarrow \infty$. The parameter, $u_{p}=\exp[-k_{0}\Delta t]=0.9$ is taken for illustration.}\label{F2}
\end{figure}

\begin{figure*}[htbp]
\includegraphics[width=0.978\textwidth]{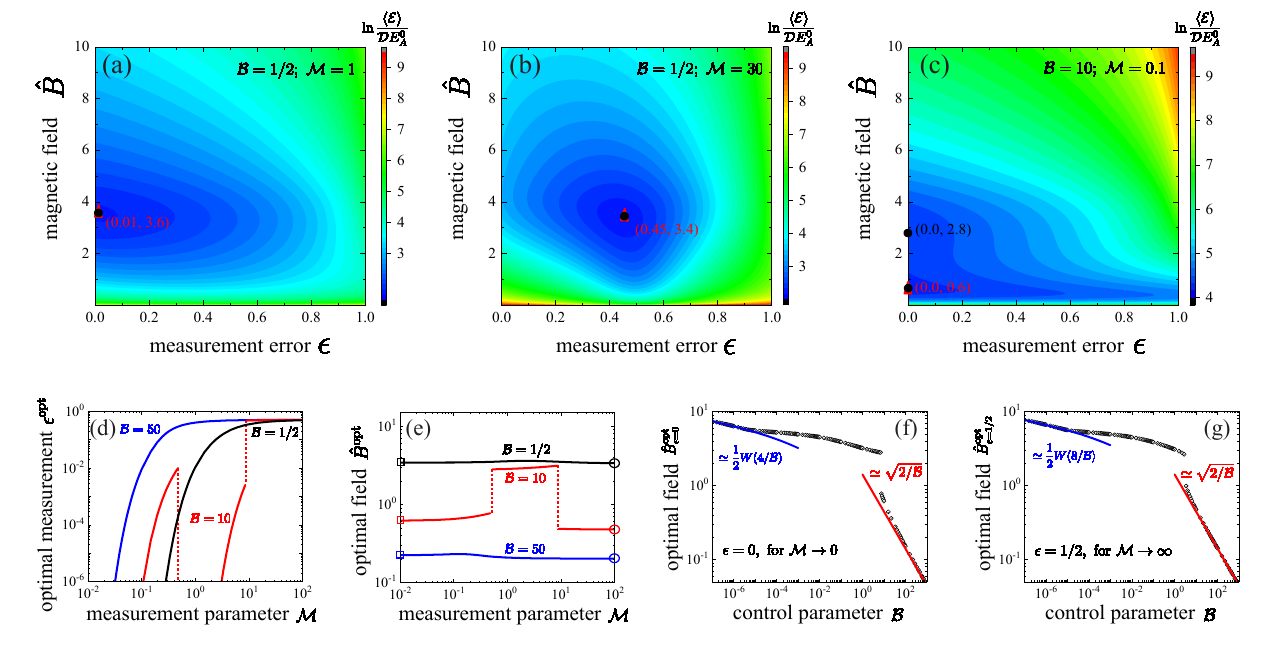}
\caption{Mean energy consumption per time step, $\ln[\langle \mathcal{E}\rangle /(\mathcal{D}E_{A}^{0})]$, as a function of the measurement error and the control field for the parameters of (a) $\mathcal{B}=1/2, \mathcal{M}=1$, (b) $\mathcal{B}=1/2$, $\mathcal{M}=30$, and (c) $\mathcal{B}=10$, $\mathcal{M}=0.1$, where the black circles denote the local minima and the red triangle denote the global minimum. The control parameter, $\mathcal{B}=E_{B}^{0}/E_{A}^{0}$, compares the magnetic-control energy scale to the active (running) energy $E_{A}^{0}$, and the measurement parameter, $\mathcal{M}=E_{M}^{0}/E_{A}^{0}$, compares the measurement energy scale to the active energy $E_{A}^{0}$. (d) Optimal measurement error and (e) optimal field as a function of the measurement parameter, $\mathcal{M}$, for the control parameter of $\mathcal{B}=1/2$ (black line), $10$ (red line) and $50$ (blue line). Squares and circles in (e) denote the values of the optimal field at the limit of $\mathcal{M}\rightarrow0$ and $\mathcal{M}\rightarrow\infty$, respectively. (f) Optimal field at the limit of $\mathcal{M}\rightarrow0 ~ (\epsilon=0)$, $\hat{B}^{\mathrm{opt}}_{\epsilon=0}$, and (g) at the limit of $\mathcal{M}\rightarrow\infty ~(\epsilon=1)$, $\hat{B}^{\mathrm{opt}}_{\epsilon=1}$, as a function of the control parameter, $\mathcal{B}$. Inset of (f) denote the values of the optimal field for the full range of $\mathcal{B}\in(10^{-6},5\times 10^{1})$.}\label{F3}
\end{figure*}

\paragraph{First passage steps -- $\mathcal{T}$.}	
Due to the stochastic nature of the system, the active particle moves along different trajectories to its destination with different first passage steps, $\mathcal{T}$, minimising the average of which can be an important regulation goal. Suppose that after $\mathcal{T}$ steps, the particle reaches the destination $x=\mathcal{D}\Delta x$.
Then the particle must move to the right with $\mathcal{W} + \mathcal{D}$ steps and to the left with $\mathcal{W}$ steps, where $\mathcal{T}=2\mathcal{W} + \mathcal{D}$.
Here, the total number of possible trajectories to reach $\mathcal{D}$ for the first time after $\mathcal{T}$ steps is $\mathcal{D}\cdot C_{\mathcal{T}}^{\mathcal{W}}/\mathcal{T}$, resembling the result of  Catalan's trapezoid problem~\cite{stembridge1997enriched}.
Correspondingly, we can obtain the  probability distribution of the first passage steps as:
\begin{align}
    P(\mathcal{T})=\frac{\mathcal{D}}{\mathcal{T}}C_{\mathcal{T}}^{\mathcal{W}}[p(n'_{k}=R)]^{\mathcal{W} + \mathcal{D}}[p(n'_{k}=L)]^{\mathcal{W}},
\end{align}
where $p(n'_{k}=R)$ and $p(n'_{k}=L)=1- p(n'_{k}=R) $ denotes the probability of the particle moving to the right and the left, respectively, and it satisfies the normalisation condition: $\sum_{\mathcal{T}=\mathcal{D}}^{\infty} P(\mathcal{T})=1$.

The average of the first passage steps is
\begin{eqnarray}
    \langle \mathcal{T}\rangle&=& \sum_{\mathcal{T}=\mathcal{D}}^{\infty}
        \mathcal{T}\cdot\frac{\mathcal{D}}{\mathcal{T}}C_{\mathcal{T}}^{\mathcal{W}}[p(n'_{k}=R)]^{\mathcal{W} + \mathcal{D}}[p(n'_{k}=L)]^{\mathcal{W}}\nonumber\\
    \quad\quad&=&\frac{\mathcal{D}}{2p(n'_{k}=R)-1},
\end{eqnarray}
which depends on the process of measurement and feedback control through the probability, $p(n'_{k}=R)$ [explicit expressions shown in the Table~\ref{t2} of the End Matter].
Note that $\langle \mathcal{T}\rangle$ diverges for unbiased random walk with $p(n'_{k}=R)=1/2$.
Apparently, the averaged first passage steps decreases with increasing magnetic field and decreasing measurement error, as shown in Fig.~\ref{F2}.
We introduce the dimensionless field strength, $\hat{B}=B/B_{0}$ with $B_{0}=k_{B}T/m$ as the referenced magnetic field.
For weak field controls, $\hat{B}\rightarrow0$,
the averaged first passage steps decreases with the increasing magnetic field in the form of $\langle \mathcal{T}\rangle/\mathcal{D} \simeq 2/[\hat{B} (1 -u_{p})]$, and for strong field controls, $\hat{B}\rightarrow \infty$, the averaged first passage steps only depends on the measurement error, in the form of $\langle \mathcal{T}\rangle/\mathcal{D} \simeq2/(1+u_{p}-2 \epsilon u_{p})$.

\paragraph{Energy consumption -- $\mathcal{E}$.}
Suppose that during each time step, $t_{k}\rightarrow t_{k+1}$, the energy consumed for sustaining the particle activity is $E_{A}^{0}$, which is usually proportional to $v_{0}^{2}$ with $v_{0}$ as self-propulsion speed, the energy of applied magnetic field is $E_{B}=E_{B}^{0}\cdot(B/B_{0})^{2}=E_{B}^{0}\hat{B}^{2}$, and the energy for measurement is $E_{M}= E_{M}^{0}\langle I(m_{k}:n_{k})\rangle=E_M^0[\ln2+\epsilon\!~\mathrm{ln}\!~\epsilon+ (1-\epsilon)\ln(1-\epsilon)]$.
Note that it is different from the case of position measurement; in that case, a particle is located at $x$ in a 1D confinement of size $L$ and its position is measured with error $\Delta x$, so the relative error is $\epsilon=\Delta x/L$ and the measurement energy is constructed simply as $E_{M}=-E_{M}^{0}\ln\epsilon$~\cite{v1_brillouin2013science,v1_Lan2012}.
Then the ensemble-averaged total energy consumed for arriving at the destination is,
\begin{eqnarray}
    \langle \mathcal{E}\rangle &=& \frac{\mathcal{D}E_{A}^{0}}{2p(n'_{k}=R)-1}~\left[1 + \frac{1}{2}\mathcal{B} \hat{B}^{2}+\mathcal{M}\langle I(m_{k}:n_{k})\rangle\right],\nonumber\\
    \label{energy}
\end{eqnarray}
where the referenced magnetic energy, $\mathcal{B}=E_{B}^{0}/E_{A}^{0}$, and referenced measurement energy, $\mathcal{M}=E_{M}^{0}/E_{A}^{0}$, are two system parameters.
Although the measurement energy is symmetric with respect to $\epsilon=1/2$, i.e., $E_{M}(\epsilon)=E_{M}(1-\epsilon)$, the symmetry gets broken when taking into account of the other two energy contributions, and minimisation of $\langle \mathcal{E}\rangle$ would always give $\epsilon\leq1/2$.

Note that the measurement energy, $E_{M}$, increases with the decreasing error for $0\leq\epsilon\leq1/2$, and meanwhile, the decreasing error can shorten the total time through biasing the probability $p(n'_{k}=R)$ and thus lowers the energy spent on the active motion and the control field; such a competition can lead to the issue of (information) robustness - (measurement) energy tradeoff, commonly discussed in biological systems~\cite{Cocconi2025,Huang2024,Ouldridge2017}.
As shown in Fig.~\ref{F3}(a), for system parameters of $\mathcal{B}=1/2$ and $\mathcal{M}=1$, the total consumed energy can have a local minimum (also the global minimum in this case) at $(\epsilon^{\mathrm{opt}}, \hat{B}^{\mathrm{opt}})$ with a very small measurement error, which serves as the optimal control variables for minimising the total energy consumption.
By increasing the measurement parameter, e.g., $\mathcal{M}=30$ in Fig.~\ref{F3}(b), there is still one local minimum serving as the global minimum, but with a relatively large measurement error, $\epsilon=0.45$.
Meanwhile, the situation can become more complicated at large $\mathcal{B}$, where there can be two local minima, as shown in Fig.~\ref{F3}(c) (see supplemental materials for more about the extrema including saddle points~\cite{supplement}).
Then the global minimum is chosen between the two local minima, which is shown in Fig.~\ref{F3}(d,e).
In this case, the system can exhibit a transition of controllable variables $(\epsilon^{\mathrm{opt}}, \hat{B}^{\mathrm{opt}})$ depending on the system parameters $(\mathcal{B}, \mathcal{M})$, where the guiding rule is to minimise the total energy consumption.
Note that the optimal field $\hat B^{\mathrm{opt}}$ appears largely insensitive to the measurement parameter $\mathcal M$, except near the strategy bifurcation points, while $\epsilon^{\mathrm{opt}}$ is shown sensitive to $\mathcal M$, and such difference roots in the different energy forms between the magnetic energy
$E_{B}=E_{B}^{0}\hat{B}^{2}$ (quadratic dependence on $\hat{B}$) and the measurement energy $E_{M}= E_{M}^{0}[\ln2+\epsilon\!~\mathrm{ln}\!~\epsilon+ (1-\epsilon)\ln(1-\epsilon)]$ (logarithmic dependence on $\epsilon$).
There are two important limits to discuss about, $\mathcal{M}=0$ [Fig.~\ref{F3}(f)] and $\mathcal{M}\rightarrow \infty$ [ Fig.~\ref{F3}(g)].
For the measurement parameter, $\mathcal{M}=0$, the measurements of optimal controls are error-free, i.e., $\epsilon^{\mathrm{opt}}=0$, and we can obtain the optimal external field follows the relation, $\hat{B}^{\mathrm{opt}}_{\epsilon=0}\simeq \frac{1}{2} W(4/\mathcal{B})$ [$W(x)$ as the principal branch of \emph{Lambert W} function] at $\mathcal{B}\rightarrow 0$, and $\hat{B}^{\mathrm{opt}}_{\epsilon=0}\simeq \sqrt{2/\mathcal{B}}$ at
$\mathcal{B}\rightarrow\infty$.
For the measurement parameter, $\mathcal{M}\rightarrow \infty$, the measurements are useless (or not performed), i.e., $\epsilon^{\mathrm{opt}}=1/2$, and we can obtain the optimal external field follows the relation, $\hat{B}^{\mathrm{opt}}_{\epsilon=1}\simeq \frac{1}{2}W(8/\mathcal{B})$ at $\mathcal{B}\rightarrow0$, and $\hat{B}^{\mathrm{opt}}_{\epsilon=1}\simeq \sqrt{2/\mathcal{B}}$ at $\mathcal{B}\rightarrow\infty$.

Moreover, the system can be generalised to become other important information-regulated active matter systems, e.g., active information engine~\cite{v1_Malgaretti2022,v1_garcia}.
If the active particle itself is regarded as a cargo, then the system setup can be applied to describe the cargo transportation process.
By introducing $\mathcal{W}=\mathcal{D}\cdot\zeta v_{0} \Delta x$ as the effective work for transporting the particle from location $x=0$ to destination $x=\mathcal{D}\Delta x$, where $\zeta$ denotes the friction coefficient of the particle and $\zeta v_{0}$ denotes the effective driving force on a passive loading,
we can define the work efficiency as $\eta=\mathcal{W}/\langle \mathcal{E}\rangle$, with $\langle \mathcal{E}\rangle$ the averaged total energy consumption introduced above.
In this simple setup of cargo transportation, the total work is constant, $\mathcal{W}=\mathcal{D}\cdot\zeta v_{0} \Delta x$, so the goal of `maximising work efficiency' is equivalent to the goal of `minimising total energy consumption', thus, the performed analyses of the minimising the total energy consumption above can be directly utilised for discussing about the work efficiency.
One can also generalise the setup to deal with more complicated information engines in the future, e.g., adding passive or actively-competing loadings on the active tracking particle and discuss about the work efficiency in different setups.

\paragraph{Summary.} In this work, we develop a \emph{smart} active model where activity, thermal fluctuation and information processing are incorporated, in analogy of a run-and-tumble bacterium equipped with a `brain'.
After discussing about the entropy production and the information flow of the system, we illustrate different optimal feedback control strategies regarding the first passage steps and the total energy consumption of the system, with a simple generalisation to active information engine.
Specifically, by setting the optimisation goal as minimising the total energy consumption, there are interesting findings, including the (information) robustness - (measurement) energy tradeoff and the strategy transition, due to the interplay between active motion, measurements and feedback controls.
Such energetics analyses can be generalised for investigating physical responses of other `smart' active systems, where the information-based regulations can be quantified to provide intuitive understandings, where one may find relevant tradeoffs of competing effects and similar strategy transitions.
Possible future connections with experiments on controllable active tracking particles are promising, e.g., platinum-nickel-gold nanorods~\cite{v1_Kline2005}, gold-polystyrene Janus particles~\cite{v1_Qian2013} and metal-dielectric Janus particles~\cite{yan2016reconfiguring,boymelgreen2022role}, however, necessary data including measurement energy, fuel consumption, control energy, etc., needs to be measured for comparisons.
Moreover, we also anticipate this study to inspire more theoretical works on formulating complex phenomena of `smart' systems, e.g., collective dynamics of information-processed systems~\cite{ Khadka2018,Xiao2024,Jung2025,Gomez-Nava2022,Sayin2025}, and also to be useful in future industrial designs of smart active systems with desired physical responses.
	
\begin{acknowledgments}	
F.M. acknowledges the supports by the National Natural Science Foundation of China (Grant No. 12275332 and 12447101), Wenzhou Institute (Grant No. WIUCASQD2023009), UCAS Xiaomi Youth Fellowship, and Beijing National Laboratory for Condensed Matter Physics (Grant No. 2023BNLCMPKF005).
\end{acknowledgments}

\newpage
\newpage
\onecolumngrid

\vspace{12pt}
\noindent\hrulefill \hspace{24pt} {\bf End Matter} \hspace{24pt} \hrulefill
\vspace{12pt}

Here we show to important probability utilised in the main text, $p(n'_{k}|n_{k},m_{k})$ and $p(n'_{k})$.
For other probabilities, one can refer to supplemental materials for their explicit expressions.~\cite{supplement}.\\

\emph{The conditional probability, $p(n'_{k}|n_{k},m_{k})$.} This denotes the transition probability for the regulatory process of $n_{k}\rightarrow n_{k}'$.

\begin{table}[htbp]
\renewcommand{\arraystretch}{1.5}
    \begin{tabular}{c||c}
        \hline
        $n_{k},m_{k},n'_{k}$ & $p(n'_{k}|n_{k},m_{k})$ \\ \hline\hline
        $L, R, L$ & $p_{\mathrm{eq}}^{0}(n_{k}'=L)+[1-p_{\mathrm{eq}}^{0}(n_{k}'=L)]\cdot e^{-k_{0}\Delta t}$ \\ \hline
        $L, R, R$ & $p_{\mathrm{eq}}^{0}(n_{k}'=R)-[1-p_{\mathrm{eq}}^{0}(n_{k}'=L)]\cdot e^{-k_{0}\Delta t}$ \\ \hline
        $R, R, L$ & $p_{\mathrm{eq}}^{0}(n_{k}'=L)-[1-p_{\mathrm{eq}}^{0}(n_{k}'=R)]\cdot e^{-k_{0}\Delta t}$ \\ \hline
        $R, R, R$ & $p_{\mathrm{eq}}^{0}(n_{k}'=R)+[1-p_{\mathrm{eq}}^{0}(n_{k}'=R)]\cdot e^{-k_{0}\Delta t}$ \\ \hline
        $L, L, L$ & $p_{\mathrm{eq}}^B(n'_{k}=L)+[1-p_{\mathrm{eq}}^B(n'_{k}=L)]\cdot e^{-k_{M}\Delta t}$ \\ \hline
        $L, L, R$ & $p_{\mathrm{eq}}^B(n'_{k}=R)-[1-p_{\mathrm{eq}}^B(n'_{k}=L)]\cdot e^{-k_{M}\Delta t}$ \\ \hline
        $R, L, L$ & $p_{\mathrm{eq}}^B(n'_{k}=L)-[1-p_{\mathrm{eq}}^B(n'_{k}=R)]\cdot e^{-k_{M}\Delta t}$ \\ \hline
        $R, L, R$ & $p_{\mathrm{eq}}^B(n'_{k}=R)+[1-p_{\mathrm{eq}}^B(n'_{k}=R)]\cdot e^{-k_{M}\Delta t}$ \\ \hline
    \end{tabular}
    \caption{The probability, $p(n'_{k}|n_{k},m_{k})$ in the regulatory process.}
    \label{t1}
\end{table}

\emph{The probability, $p(n'_{k})$.}
In the table, $p(n'_{k})$, utilised for analysing the mean first passage steps and the mean total energy consumption, depends on both the magnetic field and the measurement error explicitly.
In the table, $p_{\mathrm{eq}}^{0}(n_{k}=L, R)=p_{\mathrm{eq}}^{0}(n_{k}'=L, R)=1/2$ denotes the equilibrium distribution of $n_{k}$ and $n_{k}'$ without applying magnetic field, respectively, where the transition rate is $k_{0}=k_{0}^{L\rightarrow R}+k_{0}^{R\rightarrow L}=2k_{0}^{L\rightarrow R}$ with $k_{0}^{L\rightarrow R}$ and $k_{0}^{R\rightarrow L}$ as the transition rate from $n_{k}=L$ to $n_{k}'=R$ and that from $n_{k}=R$ to $n_{k}'=L$, respectively, without applying magnetic field.
$p_{\mathrm{eq}}^{B}(n_{k}'=L)=\exp[-mB/(k_{B}T)]/\{\exp[-mB/(k_{B}T)]+\exp[mB/(k_{B}T)]\}$ and $p_{\mathrm{eq}}^{B}(n_{k}'=R)=1-p_{\mathrm{eq}}^{B}(n_{k}'=L)$ denote the equilibrium distribution of $n_{k}'$ with applied magnetic field, and the transition rate is obtained as in the Kramers process~\cite{supplement,Kramers,hanggi1990reaction}, $k_{M}=k_{M}^{L\rightarrow R}+k_{M}^{R\rightarrow L}=k_{0}\cosh[mB/(k_{B}T)]$ with applied magnetic field.
\begin{table}[h]
\centering
\renewcommand{\arraystretch}{1.6}
\begin{tabular}{c||c}
\hline $n'_{k}$ & $p(n'_{k})$ \\ \hline\hline
\multirow{2}{*} {$L$} & $[p_{\mathrm{eq}}^{0}(n'_{k}=L)+p_{\mathrm{eq}}^{0}(n'_{k}=R)\cdot e^{-k_{0}\Delta t}]p_{\mathrm{eq}}^0(n_{k}=L) \epsilon$ \\
& $+[p_{\mathrm{eq}}^{0}(n'_{k}=L)-p_{\mathrm{eq}}^{0}(n'_{k}=L)\cdot e^{-k_{0}\Delta t}]p_{\mathrm{eq}}^{0}(n_{k}=R)(1-\epsilon)+[p_{\mathrm{eq}}^B(n'_{k}=L)+p_{\mathrm{eq}}^B(n'_{k}=R)\cdot e^{-k_{M}\Delta t}]p_{\mathrm{eq}}^{0}(n_{k}=L)(1-\epsilon)$ \\
&$+[p_{\mathrm{eq}}^B(n'_{k}=L)-p_{\mathrm{eq}}^B(n'_{k}=L)\cdot e^{-k_{M}\Delta t}]p_{\mathrm{eq}}^{0}(n_{k}=R) \epsilon$ \\ \hline
\multirow{2}{*} {$R$} & $[p_{\mathrm{eq}}^{0}(n'_{k}=R)-p_{\mathrm{eq}}^{0}(n'_{k}=R)\cdot e^{-k_{0}\Delta t}]p_{\mathrm{eq}}^0(n_{k}=L) \epsilon$ \\
& $+[p_{\mathrm{eq}}^{0}(n'_{k}=R)+p_{\mathrm{eq}}^{0}(n'_{k}=L)\cdot e^{-k_{0}\Delta t}]p_{\mathrm{eq}}^{0}(n_{k}=R)(1-\epsilon)+[p_{\mathrm{eq}}^B(n'_{k}=R)-p_{\mathrm{eq}}^B(n'_{k}=R)\cdot e^{-k_{M}\Delta t}]p_{\mathrm{eq}}^{0}(n_{k}=L)(1-\epsilon)$\\
& $+[p_{\mathrm{eq}}^B(n'_{k}=R)+p_{\mathrm{eq}}^B(n'_{k}=L)\cdot e^{-k_{M}\Delta t}]p_{\mathrm{eq}}^{0}(n_{k}=R) \epsilon$ \\ \hline
\end{tabular}
\caption{The probability, $p(n'_{k})$.}
\label{t2}
\end{table}

\end{document}